\documentclass[aps,prl,twocolumn]{revtex4}
\usepackage{amsfonts}
\usepackage{amsmath}
\usepackage{graphicx}
\usepackage{dsfont}
\usepackage{diagbox}
\usepackage{array}
\usepackage{amssymb}
\usepackage{bbm}
\usepackage{float}
\usepackage{subfigure}
\usepackage{url}
\usepackage{hyperref}
\usepackage{array}
\usepackage{multirow}
\usepackage{bm}
\usepackage{booktabs}
\usepackage{bbm}
\usepackage{array}
\usepackage{booktabs}
\usepackage{threeparttable}
\usepackage{multirow}

\newcommand{\PreserveBackslash}[1]{\let\temp=\\#1\let\\=\temp}
\newcolumntype{C}[1]{>{\PreserveBackslash\centering}p{#1}}
\newcolumntype{R}[1]{>{\PreserveBackslash\raggedleft}p{#1}}
\newcolumntype{L}[1]{>{\PreserveBackslash\raggedright}p{#1}}

\newcommand{\RNum}[1]{\uppercase\expandafter{\romannumeral #1\relax}}

\begin{document}

\title{Noise-induced entanglement transition in one-dimensional random
quantum circuits}
\author{Qi Zhang$^1$ and Guang-Ming Zhang$^{1,2}$}
\email{gmzhang@tsinghua.edu.cn}
\affiliation{$^1$State Key Laboratory of Low-Dimensional Quantum Physics and Department
of Physics, Tsinghua University, Beijing 100084, China.\\
$^2$Frontier Science Center for Quantum Information, Beijing 100084, China.}
\date{\today}

\begin{abstract}
The random quantum circuit is a minimally structured model to study the
entanglement dynamics of many-body quantum systems. In this paper, we
considered a one-dimensional quantum circuit with noisy Haar-random unitary
gates using density matrix operator and tensor contraction methods. It is
shown that the entanglement evolution of the random quantum circuits is
properly characterized by the logarithmic entanglement negativity. By
performing exact numerical calculations, we found that, as the physical
error rate is decreased below a critical value $p_c \approx 0.056$, the
logarithmic entanglement negativity changes from the area law to the volume
law, giving rise to an entanglement transition. The critical exponent of the
correlation length can be determined from the finite-size scaling analysis,
revealing the universal dynamic property of the noisy intermediate-scale
quantum devices.
\end{abstract}

\maketitle

Recently, random quantum circuits have attracted considerable attention both
theoretically and experimentally \cite%
{Google2019QuantumSupremacy,Pan2021StrongQuantum,Zhou2020WhatLimits,Nahum2018OperatorSpreading,Google2021InformationScrambling,Xu2020Accessing,Boixo2018Characterizing,Neill2019ABlueprint,Noh2020efficientclassical,Chen2021simulationNoisy,Nahum2017QuantumEntanglement,Napp2020efficient,Markov2008Simulating}%
, because it provides a minimally structured toy model to study the dynamics
of chaotic quantum many-body systems. Due to the fact that the complicated
probability distribution, it is available to achieve quantum supremacy in
sampling problems, which is impossible to simulate on classical computers at
a large scale with deep depth. The famous experimental demonstration has
been succeeded by Google's superconducting processor without any error
correction, where the device contains 53 available qubits with 20 circuit
depths\cite{Google2019QuantumSupremacy}. In a related theoretical work\cite%
{Zhou2020WhatLimits}, the quantum fidelity metrics of the random quantum
circuit has been well-studied, and they simulated Google's random quantum
circuit with tensor network states but the fidelity of two-qubit gates can
only reach $92\%$. On the other hand, the most important feature in the
dynamic process of the quantum circuits, such as the quantum entanglement,
is far less understood.

In the absence of physical noise, von Neumann entanglement entropy or R\'{e}%
nyi entropy and the corresponding entanglement spectrum are well used to
characterize the entanglement properties in quantum circuits\cite%
{Nahum2017QuantumEntanglement}. It has been established that the usual
non-unitary projective measurement may destroy the quantum entanglement in
random quantum circuits, and an entanglement phase transition is induced
from an area law phase to a volume law phase when the probability of
measurements is decreased\cite%
{Li2018QuantumZeno,Skinner2019Measurement,Bao2020Theory,Zabalo2020CriticalProperties,Choi2020QuantumError,Lunt2021Quantum,Sang2021Measurement,Lavasani2021Measurement,Li2021Statistical}%
. However, for the random quantum circuit with physical noise, it is a
challenge to separate the classical correlation from the quantum
entanglement, where the noise can thermalize the system as a mixed state.
The usual quantum mutual information for the mixed states usually
overestimates the classical correlation\cite%
{Adami1997VonNeumann,Groisman2005Quantum,Marko2008Complexity,Shapourian2021Entanglement}%
. So a natural question is how to exclusively diagnose quantum correlations
in the noisy random quantum circuits. In a related paper\cite%
{Noh2020efficientclassical}, the operator space entanglement entropy\cite%
{Zanardi2001Entanglement,Prosen2007OperatorSpace,Alba2019OperatorEntanglement}
was used to measure the entanglement of the circuit, and no entanglement
transition was observed for the gate error rate $p\geq 0.06$.

\begin{figure}[htbp]
\centering
\includegraphics[width=0.45\textwidth]{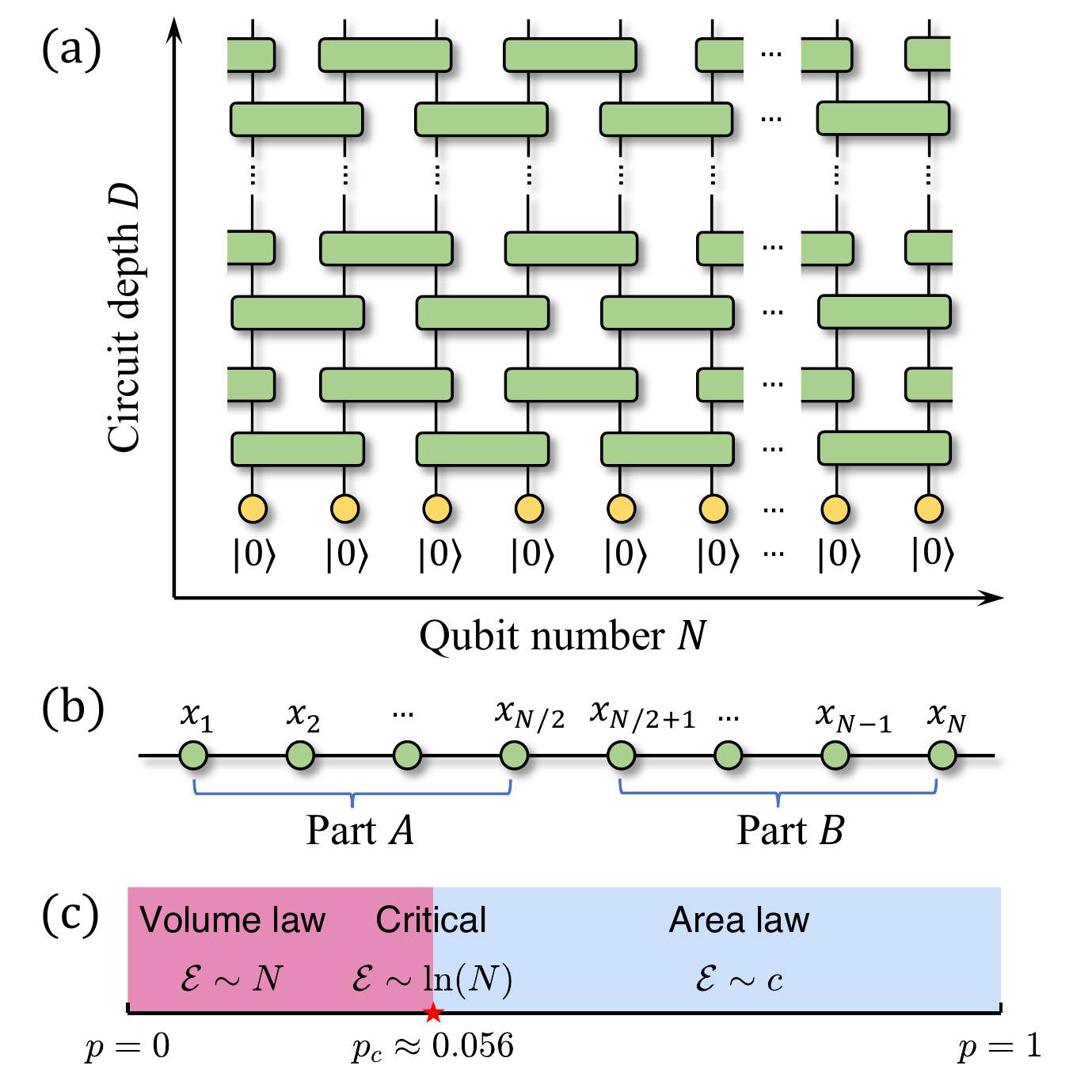}
\caption{ (a) The sketch of a one-dimensional random quantum circuit
contains $N$ qubits and circuit depth $D$ with periodic boundary conditions.
The green rectangles are two-qubit Harr-random unitary operators with a gate
error rate $p$. The yellow circles on the bottom indicate the input state,
which is a pure product state. The output distribution can be obtained on
the top of the circuit, which can be measured by the output string $%
|x_{1},x_{2},...,x_{N}\rangle $. (b) To compute the entanglement negativity,
the output mixed state is divided into two parts $A$ and $B$. (c) The
entanglement transition driven by the physical noise $p_c$ in the maximal
logarithmic negativity separates the volume law and area law. }
\label{Fig1QuantumCircuit}
\end{figure}

In this paper, we numerically simulate a one-dimensional quantum circuit
with noisy Haar-random unitary gates through tensor contractions, as shown
in Fig.\ref{Fig1QuantumCircuit}(a). We focus on the entanglement evolution
of the circuit with a lower error rate, because it is hard to accurately
simulate at a large scale with tensor networks. We notice that the
logarithmic entanglement negativity generated by the partial transpose (PT)
density matrix \cite%
{Peres1996SeparabilityCriterion,Horodecki1996Separabilityofmixedstates,Karol1998Volume,Eisert1999compareofentangle,Vidal2002ComputableMeasure,Plenio2005LogNegativity,Ruggiero2016NegspectrumCFT,Hassan2019Twistednegspectrum,Shapourian2021Entanglement,Kudlerflam2021Negativity}
can exclusively diagnose quantum correlations in noisy quantum circuits. The
corresponding bipartition is shown in Fig.\ref{Fig1QuantumCircuit}(b). By
analyzing the numerical results, we find an entanglement transition in the
maximal logarithmic negativity from the area law to the volume law as the
physical noise is decreased below a critical value $p_{c}\approx 0.056$, as
shown in Fig.\ref{Fig1QuantumCircuit}(c). We also estimate the critical
exponent of the correlation length $\nu \approx 1.25$, revealing the
universal property of the noisy intermediate-scale quantum devices.

\textit{Noisy random quantum circuit.} -Since the two-qubit gates can
approach all operations in a random quantum circuit with a finite depth, we
just consider the two-qubit gate to create the quantum correlation between
two qubits. Specially, a two-qubit Haar-random unitary gate $U$ with
dimension $2^{2}\times 2^{2}$ is applied to the system, and the unitary
operation $\mathcal{U}$ can be modeled by:
\begin{equation}
\mathcal{U}(\rho )=U\rho U^{\dagger },
\end{equation}
where $\rho $ is the density matrix of the many-body system. In a 1D quantum
circuit with $N$ qubits and circuit depth $D$, we label the qubit from $1$
to $N$, and the system starts from a pure product state, $|\Psi _{0}\rangle
=|0\rangle ^{\otimes N}$. In order to reduce the finite size effects, we use
the periodic boundary conditions in our research. When the two-qubit unitary
operations are applied, there are two kinds of patterns. For the odd circuit
depth $D$, the two-qubit gates are just applied to the qubits with labeled $%
\{l,l+1\}$, $(l=1,3,5,...)$, while the two-qubit gates are applied to the
qubits with labeled $\{l+1,l+2\}$, $(l=1,3,5,...)$ and $\{N,1\}$ for the
even circuit depth $D$. After a finite depth of two-qubit operations, the
output state can be measured by the density matrix,
\begin{equation}
\rho =\sum_{ij}q_{ij}|\Psi _{i}\rangle \langle \Psi _{j}|,
\end{equation}%
where the diagonal terms $q_{ii}$ are the probability of the basis state $%
|\Psi _{i}\rangle $. This resulting many-body state can be measured by an $N$%
-qubit string on the orthogonal basis:
\begin{equation}
|\Psi _{i}\rangle =|x_{1},x_{2},...,x_{N}\rangle \in \{|0\rangle ,|1\rangle
\}^{\otimes N}.
\end{equation}%
Then the probability of each basis state $|\Psi _{i}\rangle $ is obtained $%
q_{ii}=\langle \Psi _{i}|\rho |\Psi _{i}\rangle $, and the total probability
satisfies $\text{Tr}(\rho )=\sum_{i}q_{ii}=1$. Moreover, for a large circuit
depth, the output state converges to a pure $N$-qubit Haar-random state, the
so-called Page state\cite{Don1993Entropy,Noh2020efficientclassical}.

However, the operations on two-qubit gates cannot be completely accurate due
to the presence of several kinds of physical noise in quantum circuits. The
output state is a mixed state and is represented by the density matrix $\rho$%
. To make the study more concrete, we consider the depolarization noise,
which can be expressed as:
\begin{equation}
\mathcal{W}(\rho )=(1-p)\rho +\frac{p}{15}\sum_{W} W\rho W,
\end{equation}
where the operator $W$ belongs to the set of $15$ non-trivial two-qubit
Pauli operators $W \in \{I,X,Y,Z\}^{\otimes 2}$, and each type of noise
occurs with equal probability $p/15$. It should be noticed that the
operation $\mathcal{W}$ is not unitary because of the summation, while each
operator $W$ is also Hermitian. The operations of two-qubit gates can be
viewed as completely positive trace-preserving (CPTP) maps\cite%
{Choi1975CompletelyPositive,Noh2020efficientclassical},
\begin{equation}
\mathcal{M}(\rho )=\mathcal{U}(\rho )\circ \mathcal{W}(\rho ),
\end{equation}%
where the trace of the density matrix is preserved under the operation. The
physical meaning of each noisy Haar-random unitary gate is clear: the
quantum entanglement between two qubits is generated under the operation $%
\mathcal{U}$, while the operation $\mathcal{W}$ destroy the quantum
coherence. Hence, the quantum entanglement in such a system will first grow
up to a maximal and then decrease as the dynamic evolution. In the large
circuit depth limit, the whole system will converge to a completely and
globally depolarized state without any quantum entanglement, corresponding
to the density identity matrix.

\textit{Entanglement negativity.} -In the noisy quantum circuit, the system
under the evolution is characterized by a mixed state. To describe the
entanglement, we divide the system $|\Psi\rangle$ into two parts $A$ and $B$%
, and mainly discuss the bipartite entanglement between them. The mixed
state should be described by the density matrix $\rho$, rather than the wave
function. In the Hilbert space, the density matrix $\rho$ can be written in
an orthogonal product basis,
\begin{equation}
\rho =\sum_{ijkl}\rho_{ijkl}|\psi _{A}^{(i)},\psi _{B}^{(j)}\rangle
\langle\psi _{A}^{(k)},\psi _{B}^{(l)}|.
\end{equation}
In order to describe the mixed state entanglement, we introduce the partial
transpose (PT) density matrix,
\begin{equation}
\rho^{T_{A}}=\sum_{ijkl}\rho_{ijkl}|\psi _{A}^{(k)},\psi _{B}^{(j)}\rangle
\langle\psi _{A}^{(i)},\psi _{B}^{(l)}|,
\end{equation}
which is defined by exchanging the physical indices of part $A$.

\begin{figure*}[tbph]
\centering
\includegraphics[width=1\textwidth]{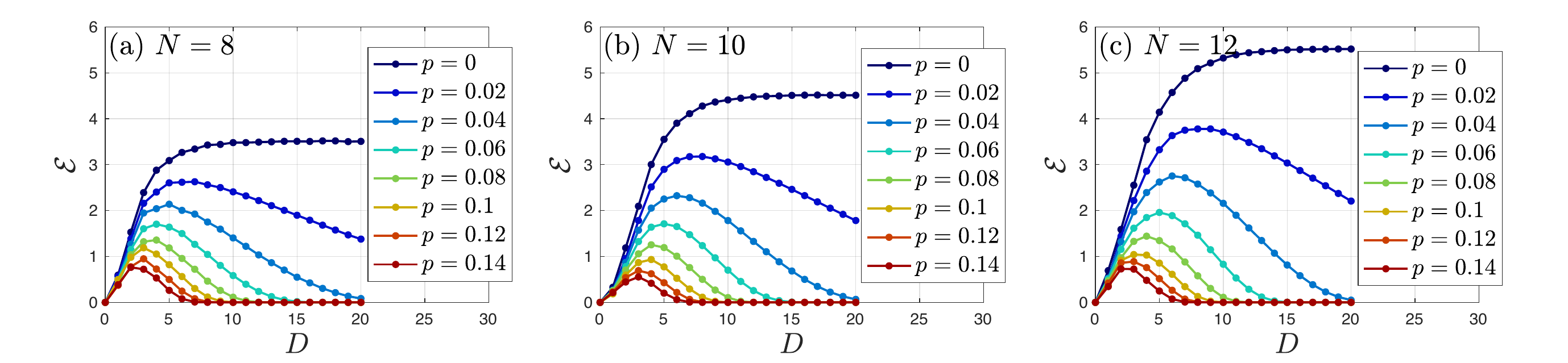}
\caption{ The logarithmic negativity $\mathcal{E}$ of the random quantum
circuit. The lines with different colors correspond to various gate error
rates $p$. The subsystem sizes of two parts are taken as $N_{A}=N_{B}=N/2$,
the total system size is taken as (a) $N=8$, (b) $N=10$, (c) $N=12$. }
\label{Fig2Variousp}
\end{figure*}

Moreover, the PT operation is a trace-preserving map, which ensures the
eigenvalues $\lambda_i$ of $\rho^{T_{A}}$ are real. Because it is not a
completely positive map, the eigenvalues of $\rho^{T_A}$ may contain
negative ones, and the existence of the negative eigenvalues is an indicator
of quantum correlations of the mixed state. The negativity can be defined by%
\cite%
{Peres1996SeparabilityCriterion,Horodecki1996Separabilityofmixedstates,Karol1998Volume,Eisert1999compareofentangle,Vidal2002ComputableMeasure,Plenio2005LogNegativity}%
:
\begin{equation}
\mathcal{N}\equiv \frac{ {\| \rho^{T_{A}} \|}_{1}-1}{2} = \sum_{\lambda_i<0}
|\lambda_i|,
\end{equation}
where $\| O \|_{1} = \text{Tr} \sqrt{O O^\dagger}$ is the trace norm, i.e.,
the sum of the absolute value of the Hermitian matrix eigenvalues. Because
the trace of the PT density matrix is preserved $\text{Tr}(\rho^{T_{A}}) = 1$%
, we can conclude that the above negativity $\mathcal{N}$ is the sum of the
absolute value of negative eigenvalues. If $\mathcal{N}>0$, the well-known
positive partial transpose (PPT) condition is violated, which indicates that
part $A$ and $B$ must be entangled\cite{Vidal2002ComputableMeasure}.
Although the PPT condition is not sufficient\cite%
{Horodecki1996Separabilityofmixedstates}, the entanglement which can not be
detected by the negativity may not be useful. So the negativity sets an
upper bound on the distillable entanglement\cite{Vidal2002ComputableMeasure}%
. For convenience, the logarithmic negativity is usually defined by:
\begin{equation}
\mathcal{E} \equiv \mathrm{log_{2}} {\| \rho^{T_{A}} \|}_{1} = \log_2 (2%
\mathcal{N}+1).
\end{equation}

Tensor network contraction is a powerful tool to simulate random quantum
circuits, especially for quantum circuits with low quantum entanglement or
shallow circuit depth\cite%
{Napp2020efficient,Zhou2020WhatLimits,Markov2008Simulating}. However, for
the high quantum entanglement system, the corresponding state cannot be
accurately described by the matrix product state or matrix product operator
with small bond dimensions\cite%
{Vidal2003EfficientClassical,VerstraeteM2004MPO,Hasting2006MPS,Verstraete2008MPS,Zwolak2004Mixedstate,Pirvu2010MPO,Zhou2020WhatLimits}%
. In the present study, the quantum circuits with physical noise ($p > 0$)
are mixed states, and they should be simulated with density matrices in the
whole space with dimension $2^{2N}$. Due to the high computational cost of
density matrix and entanglement negativity, we simulate a one-dimensional
noisy random quantum circuit chain up to $N=14$ qubits in the periodic
boundary condition. In order to compute the logarithmic entanglement
negativity of the system, the output mixed state should be divided into two
parts $A$ and $B$ with equal sizes $N_A = N_B = N/2$, as shown in Fig.\ref%
{Fig1QuantumCircuit}(b).

\textit{Numerical results.} -In Fig.\ref{Fig2Variousp}, we showed that the
logarithmic entanglement negativity $\mathcal{E}$ changes as increasing the
circuit depth $D$ for different system sizes $N$ and various gate error
rates $p$. Due to the randomness of the quantum circuit, we repeated over $%
50 $ times for the random samples and obtained the average logarithmic
negativity.

In the absence of noise $p=0$, as increasing the circuit depth $D$, the
logarithmic negativity $\mathcal{E}$ first grows linearly and then converges
to a fixed value $\mathcal{E}_{\max}$. The circuit depth $D$ corresponding
to the entanglement saturation grows with the number of qubits $N$, and the
circuit depth is proportional to the qubit number $D\sim N$, i.e., the
logarithmic negativity $\mathcal{E}$ can be regarded as a function of $D/N$.
Since the whole circuit converges to an $N$-qubit Haar-random state, the
logarithmic negativity can be expressed as $\mathcal{E}_{\max} =N/2+c_1$,
implying that the logarithmic negativity satisfied the volume law\cite%
{Shapourian2021Entanglement,Kudlerflam2021Negativity}.

In the presence of physical noise $p>0$, the logarithmic negativity $%
\mathcal{E}$ grows at first and then decreases. No matter how small the
noise is, the logarithmic negativity finally vanishes in the large circuit
depth limit. This is not surprising, because the system finally converges to
a globally depolarized state without any entanglement. In the dynamic
process of the circuit, we are most interested in the maximal achievable
entanglement and the scaling law for the maximal logarithmic negativity $%
\mathcal{E}_{\max }$.

Let us consider a small gate error rate, such as $p=0.02$. As the system
size $N$ grows, we can see the maximal logarithmic negativity $\mathcal{E}%
_{\max }$ as a linear function of the circuit depth $D$ increases, following
the entanglement volume law. But when the gate error rate is quite large,
such as $p=0.14$, it is obvious that the evolution curves for various system
sizes are almost the same, implying that the maximal logarithmic negativity $%
\mathcal{E}_{\max }$ is independent of the system size $N$, corresponding to
an area law entanglement. Hence, the scaling law of the maximal logarithmic
negativity changes with increasing the gate error rate, and there is an
entanglement transition between these two kinds of scaling law.

\begin{figure}[tbph]
\centering
\includegraphics[width=0.5\textwidth]{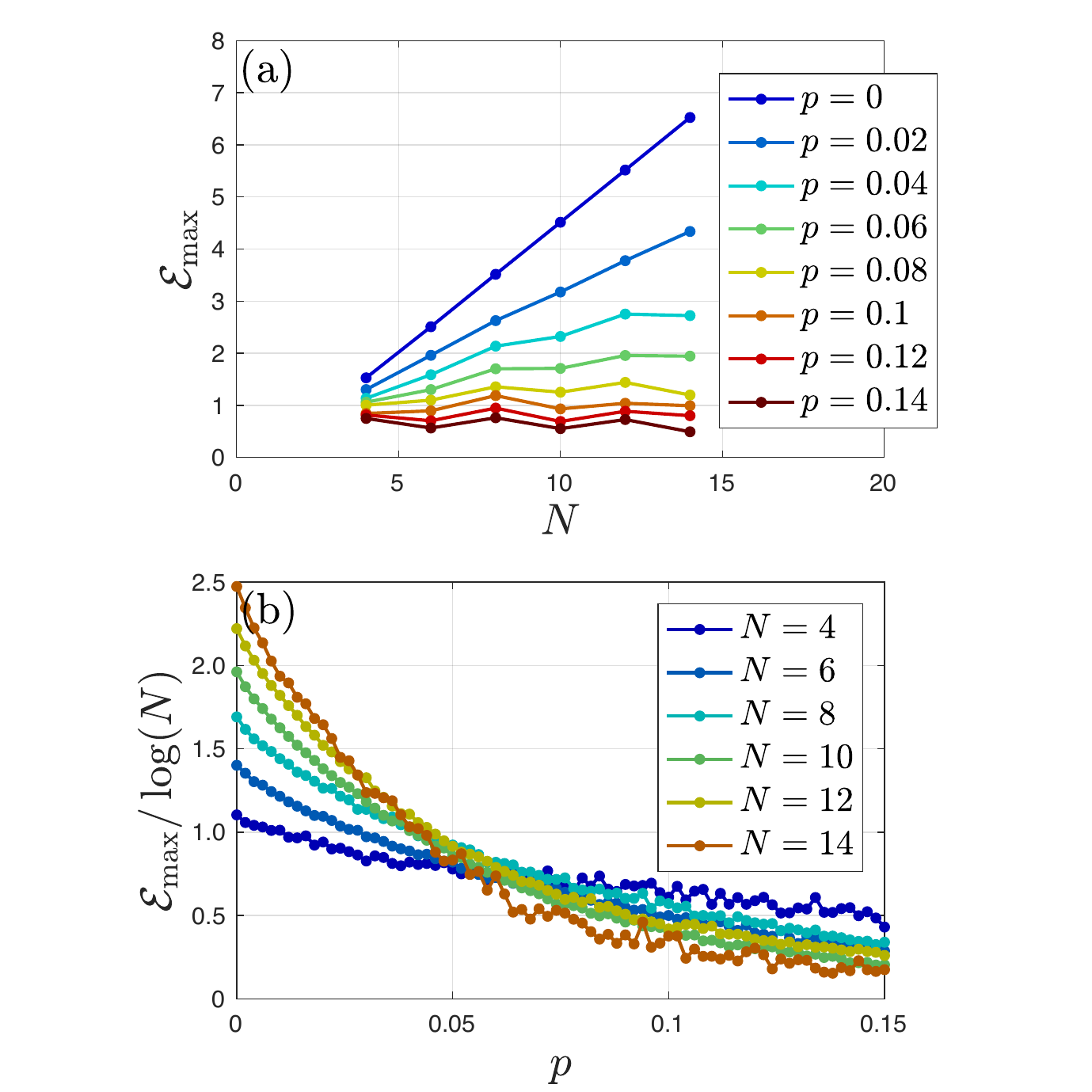}
\caption{ (a) The maximal logarithmic negativity $\mathcal{E}_{\max }$
changes as a function of the system size $N$, where all the numerical
results are taken with subsystem size $N_{A}=N_{B}=N/2$. (b) The maximal
logarithmic negativity is divided by the logarithmic of the system size $%
\mathcal{E}_{\max }/\log (N)$ versus gate error rate $p$, where the
entanglement negativities collapse together around the critical point $%
p_{c}\approx 0.056$. }
\label{Fig3ScalingLaw}
\end{figure}

\textit{Entanglement transition.} -To further analyze the scaling law of
entanglement in the noisy random quantum circuit, we choose the number of
quantum bits $N$ as the horizontal axis and the maximal logarithmic
negativity $\mathcal{E}_{\max }$ of each circuit evolution as the vertical
axis, and then connect the points with the same gate error rate $p$. The
resulting scaling law is shown in Fig.\ref{Fig3ScalingLaw}(a). It is clear
that, when the error rate $p$ is small, the maximal logarithmic negativity $%
\mathcal{E}_{\max }$ grows linearly with the system size $N$, which
corresponds to the volume law entanglement. Also, we fit the maximal
logarithmic negativity $\mathcal{E}_{\max }$ in the noiseless case $p=0$,
and then obtain the fitting formula $\mathcal{E}_{\max }=0.5001N-0.4813$,
which is consistent with the previous analytical result\cite%
{Shapourian2021Entanglement,Kudlerflam2021Negativity}. When the gate error
rate is large, such as $p=0.14$, the maximal logarithmic negativity $%
\mathcal{E}_{\max }$ does not increase with the system size $N$ growing,
which indicates the entanglement area law. Due to the finite size effect and
the odevity of subsystems $A$ and $B$, the curves slightly oscillate with
various system sizes. There should be a critical point between the area law
and the volume law entanglement, which may satisfy the $\log (N)$ correction.

In the following, when dividing the maximal logarithmic negativity by the
logarithm of the system size $\mathcal{E}_{\max }/\log (N)$, we consider its
change as a function of the gate error rate $p$. Because the numerical
results of different system sizes collapse together, we can get the
entanglement phase transition point $p_{c}$. As shown in Fig.\ref%
{Fig3ScalingLaw}(b), all the curves almost overlap at around $p_{c}\approx
0.056$. Above the critical point, the maximal logarithmic negativity
satisfies $\mathcal{E}_{\max }\sim \log (N)$. For $p<p_{c}$, the scaling law
of entanglement is stronger than the $\log (N)$ correction, entering into
the volume law phase. In contrast, when $p>p_{c}$, $\mathcal{E}_{\max }/\log
(N)$ decreases as the system size increases. Eventually, it will converge to
zero in the thermodynamic limit $N\rightarrow \infty $. Therefore, the phase
diagram can be established, as shown in Fig.\ref{Fig1QuantumCircuit}(a).

\begin{figure}[tbph]
\centering
\includegraphics[width=0.5\textwidth]{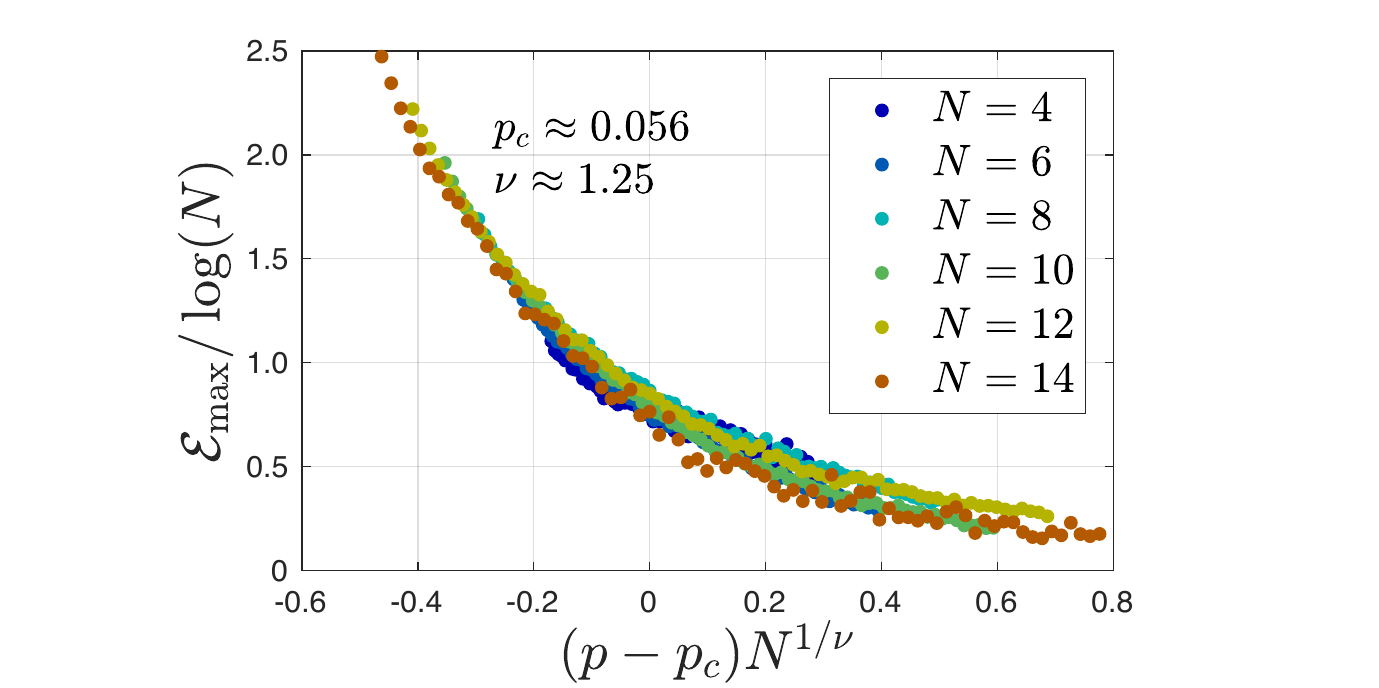}
\caption{ The maximal logarithmic negativity is divided by the logarithmic
of the system size $\mathcal{E}_{\max }/\log (N)$, which changes as a
function of $(p-p_{c})L^{1/\protect\nu }$, where the critical point is taken
as $p_{c}\approx 0.056$. The correlation length exponent $\protect\nu %
\approx 1.25$ for the best collapse. }
\label{Fig4Criticalexponent}
\end{figure}

To further study the entanglement scaling law, we can extract a critical
exponent. Since the maximal logarithmic negativity divided by the
logarithmic of the system size $\mathcal{E}_{\max }/\log (N)$ can be viewed
as a function of the gate error rate $p$ and the size of the system $N $, we
employ the following hypothesis,
\begin{equation}
\mathcal{E}_{\max }/\ln (N)\sim f\left( (p-p_{c})N^{1/\nu }\right) ,
\end{equation}%
where $f(x)$ is a universal function and $\nu $ is the correlation length
exponent. As shown in Fig.\ref{Fig4Criticalexponent}, all the points have an
excellent collapse with $\nu \approx 1.25$. On the right side of the graph,
it is the region satisfied the area law entanglement, and we find the
systems with $N=4,8,12$ and $N=6,10,14$ collapse better with each other due
to the odevity of the subsystem. It should be mentioned that both critical
point position and critical exponent are related to the entanglement measure%
\cite{Zabalo2020CriticalProperties,Li2018QuantumZeno}. It is interesting
that our numerical value is similar to the projective measure induced
transition in random quantum circuits, where through the tripartite mutual
information the correlation length exponent is $\nu \approx 1.22$ for the
circuit consisting of the Haar-random gate\cite{Zabalo2020CriticalProperties}.

\textit{Discussions and Conclusions.} -It should be noticed that the random
quantum circuits in the experiments\cite%
{Google2019QuantumSupremacy,Google2021InformationScrambling,Pan2021StrongQuantum}
have used different qubit gates, each circuit depth usually consists of one
layer fixed two-qubit gates, and one layer single-qubit gates randomly
chosen from $\sqrt{X}$, $\sqrt{Y}$, and $\sqrt{X+Y}$. Our further numerical
simulations have found that the dynamics in the circuits with two-qubit $%
\sqrt{\text{iSWAP}}$ gates are close to the Haar-random gates, where the
critical point is around $p_{c}\approx 0.040$ with the critical exponent $%
\nu \approx 1.30$.

We would also like to point out that the usual correlated many-body systems
are quite different from our random quantum circuits with high entanglement.
In most cases, the ground states satisfy the area law entanglement, which
can be described accurately by a small Hilbert space. And the large system
sizes are required to derive the scaling behavior. However, when the whole
Hilbert space is used to describe the random quantum circuits with the
volume law entanglement, the circuit with a small number of qubits has
already contained a huge Hilbert space, so the scaling behavior
can be seen even in small systems. Hence, we believe that the noise-driven
entanglement transition and its criticality found in this paper can be valid
in large-scale systems. Similarly, for the projective measurement-driven
entanglement transition\cite{Zabalo2020CriticalProperties}, the results
derived from the small size systems are comparable to the large sizes,
indicating that the entanglement phase transition is universal.

Because both the physical noise and projective measurements destroy the
quantum coherence of the system, the non-unitary projective measurement
model has some similarities to the noisy in the random quantum circuits. 
In the projective measurement-driven quantum circuit, the system is always 
in the pure states and the entanglement grows as the circuit depth increases until
saturating\cite%
{Li2018QuantumZeno,Skinner2019Measurement,Bao2020Theory,Zabalo2020CriticalProperties,Choi2020QuantumError,Lunt2021Quantum,Sang2021Measurement,Lavasani2021Measurement,Li2021Statistical}%
. To measure the mixed state entanglement in the noisy random quantum circuits, 
however, we have to use the measure of logarithmic entanglement negativity, 
which can be measured experimentally through the PT moments\cite{Elben2020MixedStateEntangleLRM}.
More recently, quantum algorithms are also proposed to measure the entanglement
negativity in noisy intermediate-scale quantum devices\cite%
{Wang2020Detecting}.

In conclusion, we have simulated the noisy random quantum circuits with
density matrix operators and tensor contractions, and characterized the
mixed state entanglement through the logarithmic entanglement negativity.
With the decreasing gate error rate, we have found that the scaling law of
the maximal logarithmic negativity changes from the area law to the volume
law. Between these two phases, the critical point has determined as $%
p_{c}\approx 0.056$ and a correlation length exponent has also been
estimated $\nu \approx 1.25$. These results not only reveal the dynamics in
chaotic quantum many-body systems, but also are valuable for designing noisy
intermediate-scale quantum devices.

\textit{Acknowledgement.} -The authors acknowledge the stimulating
discussions with Lei Wang and Tao Xiang, and the research is supported by
the National Key Research and Development Program of MOST of China
(2017YFA0302902).

\bibliography{mybibtex}

\end{document}